# The Use of Rapid Digital Game Creation to Learn Computational Thinking

**Praveen Kuruvada**
Computer Science, College of Arts and Science
Oklahoma State University, Stillwater, OK 74078, USA
praveen.kuruvada@okstate.edu

**Daniel A. Asamoah**
Information Systems, Spears School of Business
Oklahoma State University, Stillwater, OK 74078 USA
daniel.asamoah@okstate.edu

**Nikunj Dalal**
Information Systems, Spear School of Business
Oklahoma State University, Stillwater, OK 74078, USA
nik.dalal@okstate.edu

**Subhash Kak**
Computer Science, College of Arts and Science
Oklahoma State University, Stillwater, OK 74078, USA
subhash.kak@okstate.edu

**ABSTRACT**

Computational Thinking (CT) has been described as a universally applicable ability such as reading and writing. In this paper, we describe an innovative pedagogy using Rapid Digital Game Creation (RDGC) for learning CT skills. RDGC involves the rapid building of digital games with high-level software that requires little or no programming knowledge. We analyze how RDGC supports various CT concepts and how it may be mapped to equivalent Java concepts by building the same game using both RDGC and Java. We discuss the potential benefits of this approach for attracting computing majors, as a precursor to learning formal programming languages, for learning domain knowledge, and for bridging the digital divide. We present the implications of this work for teachers and researchers.

**INTRODUCTION**

Computational Thinking (CT) is an important type of thinking, which combines key elements of analytical, critical, and creative thinking. Wing's (2006) seminal article on CT states that "computational thinking represents a universally applicable attitude and skill set that everyone, not just computer scientists, would be eager to learn and use." CT is concerned with conceptualizing, problem-solving and designing systems drawing upon mathematical and engineering thinking using concepts fundamental to computing (Wing, 2006).

However, the mode and method for teaching CT still remains a challenge even though it may be seen as a fundamental skill for problem solving in all disciplines (Guzdial, 2008). To this end, the use of Rapid Digital Game Creation (RDGC) has been proposed as a pedagogical framework for teaching CT in an innovative way (Dalal et. al, 2009a) because video games are attractive and captivating to all groups of people including both adults and children of both genders. RDGC is the process used to build computer games quickly and easily using game creation software that requires little or no programming knowledge. Rapid game creation enables a creator to build a quick prototype game and to see the effects of changes almost immediately (Dalal, et. al, 2009b, p. 125)." Curricula that have used game design in teaching computing concepts have largely found positive effects on students (e.g., Bayliss and Strout, 2006; Parberry, Kazemzadeh and Roden, 2006).

**In this paper, we describe how RDGC supports the learning of basic and advanced concepts in CT.** We show how RDGC learning can be mapped to equivalent concepts in a formal programming language. This paper is organized as follows. In the next section, we describe RDGC and its benefits in teaching CT skills. Next, we discuss how RDGC supports the learning of CT. Using a rapid game development tool called Game Maker, we demonstrate how RDGC could be used as a pedagogical framework to teach CT. In particular, we demonstrate the building of a Pong game and how that can be used to learn CT concepts necessary for computer programming. We map the RDGC implementation constructs to equivalent concepts in the popular Java programming language. We conclude with a discussion on the effectiveness of RDGC in learning CT and discuss implications for future teaching and research.

**RAPID DIGITAL GAME CREATION**

As described earlier, Rapid Computer Game Creation refers to the process of building computer games quickly and easily, using game creation software that requires little or no programming knowledge. RDGC offers an easy and enjoyable way of achieving this task of building computer games. It does not require the user to have prior mastery over all the fundamental constructs of programming. RDGC uses an Object Oriented platform to build templates and offer options for choosing objects, events and methods. Hence, in an attempt to build a video game, users intrinsically learn basic programming concepts without necessarily realizing that they are using those concepts. Subsequently, when they do learn programming, it is easier for them to understand the programming constructs because they can be correlated with specific examples from the user's own game products.

There are various RDGC tools available such as Game Maker (http://www.yoyogames.com/), Multimedia Fusion (http://www.clickteam.com/website/usa/), Alice (http://www.alice.org/), and Scratch (http://scratch.mit.edu/). They vary in several aspects such as ease of use and in the availability of different complex options for building a game. For example, Alice is a 3D programming environment used for creating animations for story telling or games (Conway, 1997). Scratch is also a similar game creation tool used for creating animations, music and art. **The concepts described in this paper are platform independent** as any of the tools can be used as a means to enhance CT. For the purpose of this paper, we have chosen Game Maker as the platform because it is popularly available in the public domain and because of the relatively short learning curve it requires (Habgood and Overmars, 2006). Besides, Game Maker has got several options such as Sprites (where the user gets the options to create the characters required for the game being designed), Objects (where the characters created in sprites can be imported and assigned with some events), Room (the actual window of the game) and others.

**Designing A Game In Game Maker**

In this section, we take a look at some of the key components to consider in creating a video game using the Game Maker software. We take a game template we designed using Game Maker and draw parallels with the equivalent game designed in Java and subsequently show how it supports CT concepts. We explore how CT as needed for proficiency in a higher level language like Java can be taught to novices in computer programming using the easy-to-use features of a Rapid Computer Game Creation Tool as a precursor to learning the formal language.

Figure 1 shows a screen shot from a prototype Pong game created using Game Maker. The Pong game was designed as a template to illustrate the various programming concepts. The time it takes to create the initial game is as little as 30 minutes. An equivalent Pong game was created using the Java programming language in an effort to understand the correspondence between an RDGC tool and a formal programming language.

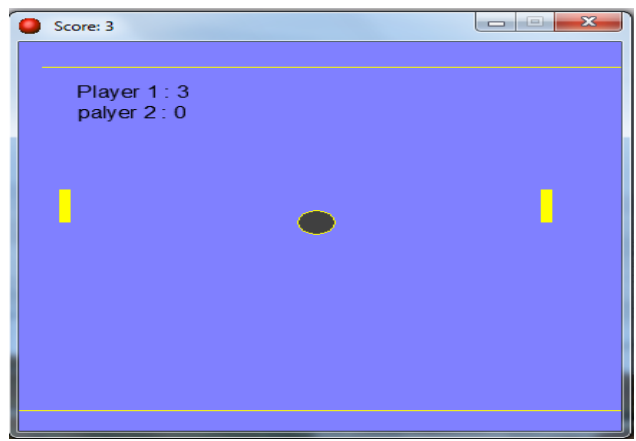

**Figure 1. Prototype Pong Game designed using GameMaker**

Designing the game in Game Maker requires very little or no programming knowledge. It has several in-built menu options that can be selected as per the user's requirements. Initially we need to create or upload the characters required for the game. This is done in the sprites section where we have an interface where the user could draw a character or upload an existing character.

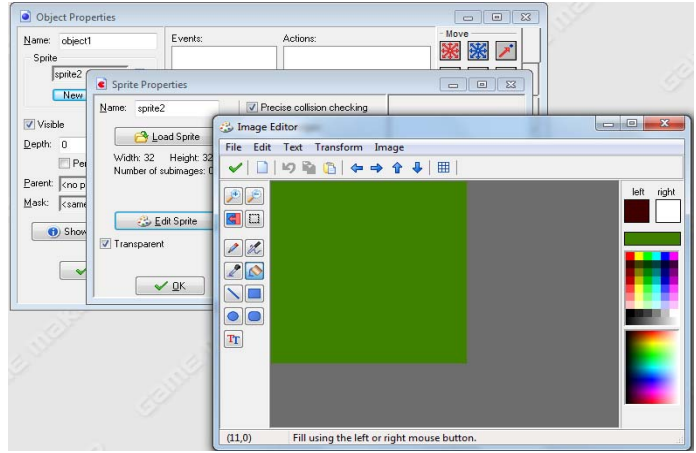

**Figure 2. Designing an object**

Once the required sprites are created, we import them to the objects. Once the objects are created, the next step is to assign events to the objects so that they perform actions as required by the game (see figures 2 and 3). The use of 'Events' helps the user iterate through different 'if-statements' in trying to make choices as required for the games. At this stage, the user learns what sort of action would be performed upon selection of a particular option. It forms a basis to better understand the basic concepts of programming such as the use of sequence, loops, decision structures, and other aspects of programming. The user gets a better understanding of what event would be performed when a particular option is selected.

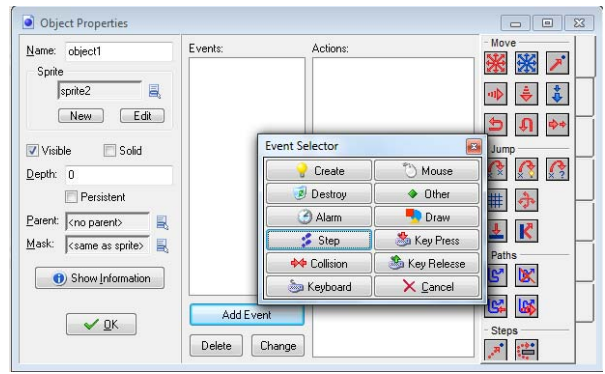

**Figure 3. Adding events to objects in Game Maker**

After assigning the required events and setting up other aspects of the game such as the score board and the room design where the objects are to be placed (see figure 4), the final game is ready to be played.

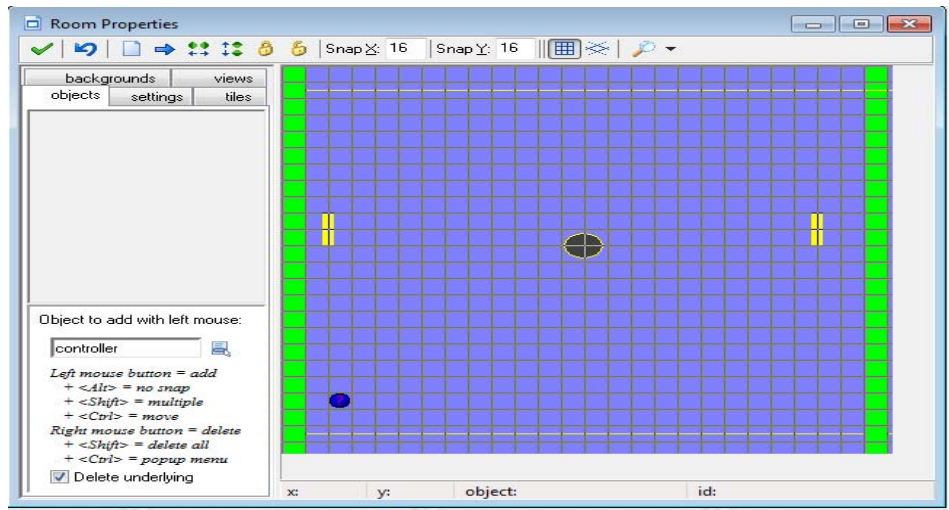

**Figure 4. Design of room where the objects are to be placed**

**RDGC AND COMPUTATIONAL THINKING**

How does designing the game using an RDGC tool help improve CT? To understand this issue, we designed an equivalent Pong game in Java and mapped the explicit programming concepts to various aspects of RDGC. We believe that the use of RDGC by the user before being introduced to formal programming would create a better understanding and improve programming skills (Dalal, Dalal and Kak, 2009). Through RDGC, concepts that relate to CT such as object-orientation, procedures, abstraction and problem decomposition can be learned (Wing, 2006).

A key aspect of CT involves object-oriented thinking and an understanding of concepts such as objects, events, abstraction, polymorphism, encapsulation among others. CT also involves the understanding of programming structures such as sequence, decisions, and iterations. Table 1 shows some programming concepts involved in languages such as Java.

| Concept level | Concept | Description |
|---|---|---|
| Higher-level concepts | 1) Abstraction | Abstraction is used to represent essential characteristics without necessarily explaining all the details. |
| | 2) Inheritance | Inheritance allows an object of a class to acquire the properties of the object of another class |
| | 3) Polymorphism | Polymorphism allows an operation to take more than one form and hence, show different behaviors in different situations |
| | 4) Encapsulation | Encapsulation compartmentalizes the functional details of some or all of the object's components such that access is restricted within that particular object. |
| Detailed concepts | 1) Java Swings | Java Swings are used to create a GUI or an interface where the objects that perform specific actions can be added. |
| | 2) Graphics | Graphics are used to draw and paint objects in the window |
| | 3) Events | Events determine the flow actions in the program |
| | 4) Loops | Loops are used to perform an action or a set of instructions continuously until a condition occurs |
| | 5) Others | Constructors are used to initialize variables as the object is created. |

**Table 1. Some Java concepts that could be learned through RDGC**

All these concepts show how game designing would present a creative approach to learning basic programming concepts. All the concepts and their implementation in the design of the game are explained below.

**Basic Concepts**

*Inheritance*

Object-oriented programming allows classes to *inherit* commonly used state and behavior from other classes. Every class inherits object classes in the program by default. The window class inherits the default applet class and the play class inherits JFrames in the game designed using Java.

This method of use of inheritance could be compared to some options in Game Maker which appear similar in action to that of inheritance though they logically do not mean the same. Creation of sprites and inheriting the sprites or their properties and creating objects can be taken as an example for demonstrating inheritance in Game Maker. This is shown in table 2.

| Concept | Sub concepts | Description | RDGC implementation | Java Code |
|---|---|---|---|---|
| Inheritance | 1) Single Inheritance<br><br>2) Multiple inheritance<br><br>3) Multilevel Inheritance<br><br>4) Hierarchical inheritance | Inheritance is the capability of a class to use the properties and methods of another class while adding its own functionality | 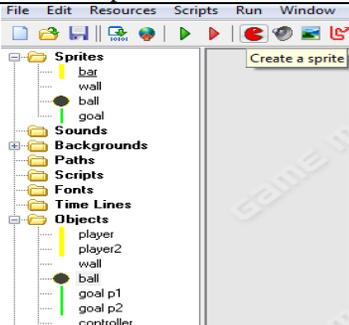<br><br>An example of the use of inheritance concept in game maker would be the use of objects inheriting the properties of sprites in the pong game. | Equivalent Java implementations showing how classes inherit properties form other.<br><br>Class bar extends Object<br>{      ……………<br>}<br>Where object is the default in-built class.<br>EX:<br>Class window extends JPanel<br>{      ………….<br>} |

**Table 2. Inheritance concepts in Java**

*Polymorphism*

Polymorphism is a programming language concept that allows values of different data types to be handled using a uniform interface. Polymorphism concepts such as operator overloading are used in the program.

| Concept | Sub concepts | Description | RDGC implementation | Java Code |
|---|---|---|---|---|
| Polymorphism | 1) Run Time polymorphism:<br>  a) virtual function<br><br>2) Compile Time polymorphism<br>  a) function overloading<br>  b) operator overloading | Polymorphism is a programming language feature that allows values of different data types to be handled using a uniform interface. | The two bars in the pong game can be considered as an example for polymorphism since both bars in the game use the same sprite for their creation but perform different actions in the game. | Code shows same method being called using different parameter list.<br><br>Public void setPosition (Point position)<br>{      …….}<br>public void setPosition(double xps, double yps)<br>{      …… } |

**Table 3. Types of polymorphism**

*Encapsulation*

Encapsulation is a language mechanism for restricting access to some of the object's components. By using access specifiers like public, private and protected, we can restrict access to required methods.

When declared as a public method, the update method can be called by objects in other classes as shown below.
    public void update ()

```
        {    …..
        }
```

Also, making the randomized method private will restrict the use of the method to the class in which it is defined as shown below.

```
        private void randomizeBall ()
        {    …….
        }
```

**Detailed Concepts**

*Use of Java swings and applets*

Java swings are used to create the window where the game is to be played and to add the applet components in the game.

    JFrame f1=new JFrame("game");

    getContentPane().setLayout(null);

    setSize(width + 8, height + 8);

The creation of a room in Game Maker, placing various objects in the room and setting up the dimensions of the room can be compared to that of using the swings and applets in Java to create a window for the game.

*Graphics*

This component in Java is used to create shapes required in the game, paint them and do appropriate changes as required in the game.

Creation of sprites, coloring them, setting their dimensions and various other options in Game Maker as shown in table 4 can be compared to that of using graphics in Java for designing.

| Concept | Sub concepts | Description | RDGC implementation | Java Code |
|---|---|---|---|---|
| Graphics | 1)2D graphic API<br><br>2)3D graphic API | There are many options in Java to enable graphics. Some examples are AWT (applet window toolkit) and SWINGS. When the Java applet is activated, Java looks for a method called **Paint** which has a single parameter of type, **Graphics** | 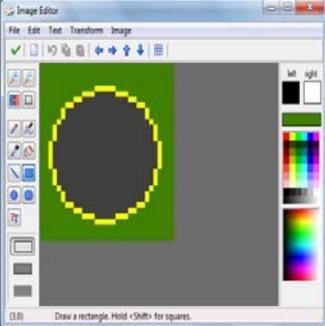<br><br>Graphics concept is used to draw the characters required for the game. The design process requires understanding of dimensions, alignment etc. | public void paint(Graphics g)<br>{<br>  g.setColor(new Color(0,0,0));<br>  g.fillOval(x,y,w,h,);<br>}<br>This will draw t hovel shaped ball<br><br>public void paint(Graphics g)<br>{<br>g.setColor(new Color(0,0,0,0));<br>bll.paint(g);<br>player1.paint(g);<br>player2.paint(g);<br>}<br>This would paint the object created. |

**Table 4. Use of graphics in Java and relevant ways of creating graphical objects in Game Maker**

*Event*

Events refer to the significant occurrence or something that takes place upon an action performed by the user. There are various types of events in Java such as window events, key events, and mouse events.

We do have similar options while assigning events to the objects created in Game Maker. These events allow the user to decide the specific task that is to be performed by the object upon assignment. Adding events to the objects and immediately observing the changes would provide the user a better understanding of the events as used in programming.

| Concept | Sub concepts | Description | RDGC implementation | Java Code |
|---|---|---|---|---|
| Event Handling | 1) Action Event<br>2) Adjustment Event<br>3) Component Event<br>4) Container Event<br>5) Focus Event<br>6) Item Event<br>7) Key Event<br>8) Mouse Event<br>9) Paint Event<br>10) Window Event | Event is an action that is usually initiated outside the scope of a program and that is handled by a piece of code inside the program | 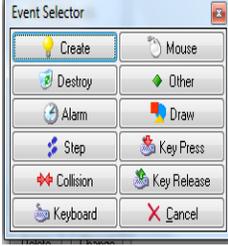<br>The Event Menu shown above can be considered as an example for the use of different types of events while building a game using game maker. | aTimer = new Timer(10, new TimerRepaintListener());<br>This event generates a time interval of 10 ms for each update to be performed.<br><br>this.addKeyListener (new barListener());<br>This event moves the bars in the pong game by listening to the user actions. This is performed using the key events. |

**Table 5. Types of events in Java and ways to assign events to objects in Game Maker**

*Loops and Constructors*

The use of basic loop structures such as the If–else structure in the design of the games makes a program perform a specific action or instruction repeatedly till a condition occurs. Examples of loops include while loop, do-while loop, for loop, if-else ladder and else-if ladder.

Use of various events to perform an action repeatedly can be compared to that of loops in a programming language.

A constructor has the same name as that of the class name and is used to initialize the variables in the class as the object is created. It is defined with the same name as the class without any return type.

| Concept | Sub concepts | Description | RDGC implementation | Code |
|---|---|---|---|---|
| Loops | 1) For loop<br>2) While loop<br>3) Do while loop<br>4) If else loops | A loop is a control flow statement that allows code to be executed repeatedly based on a boolean condition. | During the process of building a game in Game Maker, the user goes through many if-else statements. For example, the user has to decide when a game would terminate or when a player wins. Following this process while building a game will make him better understand loops and control flows. | if (Condition)<br>{ action to be performed }<br>else<br>{ action to be performed} |

**Table 6. Types of loops in Java and similar application in Game Maker**

**DISCUSSION**

In this paper, we have introduced the notion of Rapid Computer Game Creation and discussed ways in which RDGC supports CT. We do *not* suggest that game creation is better with RDGC as compared to Java. Rather, as we have argued, when RDGC is used as a precursor to teaching a formal programming language such as Java, there is potential for the student to subsequently gain a better understanding of the programming constructs if the instructor explains the constructs in terms of the game that the student has created. Hence, the two approaches – RDGC and Java (or any other formal programming language) complement one another.  We believe that game building, as a pedagogical model helps teach CT in a dynamic manner which in turn flattens the steep learning curve needed to learn computer programming.

There is some evidence that game building, as a pedagogical model would be attractive to students even if they are not from the onset interested in learning about CT concepts (Moreno-Ger, Burgos, Martinez-Ortiz, Sierra and Fernandez-Manjon, 2008).  RDGC offers a relatively quick and easy way of teaching basic CT skills. We believe that the CT skills gained through the RDGC tool is an effective way of teaching rudiments of CT to all majors. In addition, RDGC can be used to teach domain-specific knowledge if students are required to build games in specific domains. The pedagogical advantages in other domains of study stem from the fact that CT offers a fundamental idea for formulating viable questions and encourages thinking through possible answers to them. This approach to solving problems goes beyond Creative Thinking and Critical Thinking.

The Pong game discussed in this paper was built using both Game Maker and Java programming language. It should be noted that the time taken to design and build the game was significantly higher in Java than with the RDGC tool. The **rapid** nature of RDGC and the relative ease of learning and use of the tool permit the learning of CT and problem-solving in multiple contexts and at early stages of a student's college career. Easier learning would help alleviate some of the problems that have led to the downward trend in enrollment in Computer Science (CS) and Information Systems (IS) disciplines in recent years. This approach to learning CT is helpful, especially when there is a general sense among students that subject content areas such as computer programming are cumbersome, unexciting or monotonous. RDGC has the potential to teach CT and enable the understanding of basic programming concepts in a fun way. Fostering good CT skills in students at an early stage will decrease their fear for disciplines that require some aptitude in computer programming. If CT fundamentals are taught at the pre-college or early college levels, students would feel more confident in choosing majors in the computing field. This is even more important given that the current and revised model curricula for IS education has strongly recognized the need for computing professionals who have strong analytical and critical thinking skills (Topi, Valacich, Kaiser, Nunamaker, Sipior, Vreede and Wright, 2009).

Moreover, the gap between those who use information and computer technology and those who do not has widened in recent years. To reduce this gap, Rapid Computer Game Creation (RDGC) may be used as a way of introducing students to CT and subsequently bridging the digital divide (Dalal, Dalal, Kak, Antonenko and Stansberry, 2009). Statistically, women, under-represented minorities and the elderly are some of the demographic groups that are on the low end of the digital divide. Since the appeal of games transcends gender, age, and race, introducing RDGC can potentially increase CS and IS enrollment among groups historically known to be under-represented in those disciplines.

Learning CT has implications not only for skills in computing but also for other fields of study and in daily living. After all, CT skills help us to think abstractly. For instance, learning basic algebra or other fundamental quantitative skill does not necessarily imply a career path in higher level mathematics (Lu and Fletcher, 2009). Rather, knowledge in quantitative skills can be extended into the visual arts and even music. Thinking in a computational manner is essential to decision making in daily living. Inadvertently, we daily iterate through different 'if-else statements' in trying to make choices in different situations. This inert human capability can be enhanced through the teaching of CT skills to students. Hence, besides computer programming, learners think through uncertainties and sift through viable options out of a load of information (Wing, 2006).

RDGC may have other benefits besides learning CT skills. According to Richard Ferdig and Jeff Boyer (2007), students tend to develop positive self esteem and sense of accomplishment if they are able to create their own video games by use of the RDGC tools. Also, when students utilize game creation to develop a game in a particular academic domain, they develop a deeper insight into that particular domain area. Besides, when students combine content across a spectrum of offerings in the curriculum, they broaden their general understanding of their field. For instance, in physics, game templates and simulations can be built to learn or teach specific concepts. Research has found that higher proficiency and understanding of mathematical concepts could be developed by learning computer programming (Papanastasiou and Ferdig, 2006).

The work reported in this paper raises a number of research issues for pedagogy on aspects related to RDGC. These have implications for teachers and researchers. There is need for empirical studies to understand how well the user can learn CT concepts using RDGC. There is also a need for effective pedagogic models and best practices for the use of RDGC in the classroom. Theoretical foundations for CT and RDGC need to be investigated. Other issues that emerge include research into the use of pre-built template games for imparting domain-specific knowledge and CT skills. We would need to explore the types of games that appeal to different kinds of users in order to facilitate the building of an effective RDGC pedagogic framework.

## CONCLUSION

CS and IS education needs more innovative ways of instruction. Rapid Computer Game Creation has the potential to be an effective pedagogical model in IS and computing courses. Although a lot more research is needed in this area, because of the wide popularity of digital gaming, RDGC holds a lot of promise as a way to teach CT in any computing course and as a precursor to programming courses. It is also a useful pedagogic tool for other academic areas and not just content areas that require computer programming. Game construction and game playing provides more flexibility since it uses a variety of objects and scenarios in an interactive environment. Curriculum designers must consider the inclusion of CT at the pre-college level at par with other fundamental skills such as reading, writing and algebra. Providing students with pre-designed games templates and guiding them to build computer games rapidly constitutes a creative approach for promoting CT and increasing interest in computing disciplines.